\documentclass[onecolumn, amsmath,amssymb, mathtools, prf]{revtex4-2}

\usepackage{graphicx}% Include figure files
\usepackage{dcolumn}% Align table columns on decimal point
\usepackage{mathtools}% Align table columns on decimal point
\usepackage{bm}% bold math
\usepackage{cancel}
\usepackage{nicefrac}
\usepackage{xcolor}
\usepackage{bigints}
\usepackage{tikz}
\usepackage{cleveref}
\usepackage{bbm}
\usepackage{subcaption}
\usepackage{relsize}
\usepackage{amssymb}
\usepackage{amsmath}
\crefname{figure}{Fig.}{Figs.}
\crefname{equation}{Eq.}{Eqs.}
\crefformat{equation}{Eq.~#2#1#3}

\begin{document}

\title{Effective Viscosity of a Suspension of Hot Particles}

\author{Osher Arbib}
 \altaffiliation{School of Mathematical Sciences, Tel Aviv University, Tel Aviv 6997801, Israel.}
  \email{osherarbib@mail.tau.ac.il}
\author{Naomi Oppenheimer}%
 \altaffiliation{School of Physics and Astronomy and the Center for Physics and Chemistry of Living Systems, Tel Aviv University, Tel Aviv 6997801, Israel.}

\date{\today}

\begin{abstract}
Active particles with a temperature distribution, ``hot particles", have a distinct effect on the fluid that surrounds them. The temperature gradients they create deem the fluid's viscosity spatially dependent, therefore violating the linearity of the problem, making a full description of velocity and pressure fields challenging. Using energy dissipation analysis and Lorentz Reciprocal Theorem, we show that it is still possible to study global properties of such hot suspensions. 
Namely, we calculate the effective stress tensor and effective viscosity of a dilute hot suspension, adding a correction that includes contributions from the bulk fluid and the particles themselves. As examples of this method, we derive the effective stress and viscosity of a suspension of spherical particles with different heat distributions. We show that when the particles are non-Brownian and are all oriented along the same direction, the viscosity is no longer isotropic and depends on the direction of shear relative to the orientation of the particles. In such cases, the fluid is also non-Newtonian. If the particles' orientation is fixed due to an external field, the stress tensor is no longer symmetric, and the viscosity has odd components. 

\end{abstract}

\maketitle
\section{Introduction}

The viscosity, $\eta$, of a fluid is a measure of its resistance to flow. 
A common way to measure the viscosity is to place a fluid between two walls, 
hold the bottom fixed and move the upper surface with a constant velocity. 
The force density required to move the 
upper surface with respect to the bottom one is a measure of the viscosity. 
When a bare fluid undergoes shear, it experiences a certain stress. If the fluid is not bare but includes a volume fraction $\varphi$ of particles, the viscosity increases. In his famous work, Einstein showed that, to a leading order, the effective viscosity is given by \cite{Ein1906}
\begin{equation}\label{eq:Eins_rel}
    \eta^\mathrm{eff}=\eta^{\infty}\left(1+\frac{5}{2}\varphi\right),
\end{equation}
where $\eta^{\infty}$ is the viscosity of the bare fluid (throughout this work, we use superscripts to identify the setting --- e.g., bare fluid, particulate fluid, hot particle --- subscripts are used as indices). 
Assuming particles in the solution are force and torque-free, the leading order contribution they have on the flow is through the stresslet, which is the symmetric part of the force-dipole.
Specifically, for a rigid spherical particle with a radius $a$, subject to a constant shear rate, ${\bf E}^{\infty}$, the stresslet is given by \cite{leal07}
\begin{equation}\label{eq:stresslet_sphere}
    \boldsymbol{\mathcal{S}} = \frac{20\pi a^3\eta^{\infty}}{3}{\bf E^{\infty}}.
\end{equation}
The presence of particles in the solution alters the stress through their stresslets. The new ratio between stress and strain rate is a possible way of deriving \cref{eq:Eins_rel} --- the average contribution of the particles is given by multiplying \cref{eq:stresslet_sphere} by the number of particles and dividing by the total volume \cite{batchelor1970stress}. %Dividing Eq.~\ref{eq:stresslet_sphere} by the volume of the particle and XXX gives  
An alternative derivation of the average viscosity was given by Brenner \cite{brenner58, brenner1970rheology}, who showed that the effective viscosity could be defined as the ratio of the energy dissipation of the particulate fluid, denoted as $\Phi$, to that of the bare fluid, given by $\Phi^{\infty}$, 
\begin{equation}\label{eq:eff_vis_brenner}
    \eta^\mathrm{eff}=\eta^{\infty}\frac{\Phi}{\Phi^{\infty}}.
\end{equation}

Both methods have been used to extend Einstein's relation and calculate the effective viscosity of various settings, such as cases where the particles are not spherical~\cite{jeffery1922motion, brenner1972suspension}, non-rigid particles (drops)~\cite{Taylor1932}, emulsions in a non-Newtonian fluid \cite{oldroyd1953elastic,danov2001viscosity, chhabra2006bubbles}, semi-dilute suspensions \cite{batchelor1972determination} which include higher order terms in the volume fraction.
The effective viscosity of active particles, i.e., particles that propel themselves, such as bacteria, was also calculated~\cite{haines2012effective}.
Recently, the effective viscosity of a suspension of slip-stick particles was derived \cite{janus22}.

For a bare-fluid, the viscosity can be constant or spatially dependent, which can come from environmental properties such as external fields (e.g magnetic field), temperature gradient or the existence of active particles and achiral molecules \cite{khain2022stokes, everts2023dissipative, eastham2020axisymmetric, shaik2021hydrodynamics}.  
In this paper, we focus on a case in which the particles immersed in the fluid actively change its viscosity. A common example is of a colloidal suspension that is heated by an external field such that the temperature distribution around each particle is spatially varying, $T = T({\bf r})$ \cite{Jiang2010, moyses2016trochoidal, das2022flowfielddisturbancepoint}. Since viscosity is strongly temperature dependent, the temperature field will induce a spatially varying viscosity $\eta({\bf r})$. A hot colloidal particle will have altered Broniwan motion \cite{kroy2016hot, Rings2010, Falasco2014, Chakraborty2011}. The presence of temperature gradients could cause particle motion due to thermophoresis \cite{piazza2008thermophoresis,wurger2007thermophoresis}, or other interesting dynamics may emerge~\cite{Schermer2011,holubec2020active}.   The ability to control externally the effective viscosity can be useful and can be applied in 
various cases such as drug delivery, heat transfer, or for nanomotors (see, e.g., \cite{patel2012effective}).

\begin{figure}[h]
    \centering

    {\includegraphics[width=.1\textwidth]{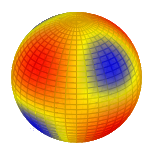}}

\begin{tikzpicture}[thick,scale=0.2, every node/.style={scale=2.6, align=left}]

  \draw[-, line width=2](16,5.5)--(18,5.5);
  \draw[-, line width=2](16,4.5)--(18,4.5);
       \fill[fill=green, fill opacity=1 , ball color=green, shading=ball] (-3,0.0) circle(2);
    \draw[-, line width=2](0,0)--(2,0);
  \draw[-, line width=2](1,1)--(1,-1);
  \draw (5, 0) node(formula){$\xi\Bigg($};

       \fill[fill=red, fill opacity=1 , ball color=red, shading=ball] (8,0.0) circle(2);
  % \draw node[anchor=west](12, 0) \text[fontscale=3]{%Stokes' Law\\ 
  %   $+$
  %   };
    \draw[-, line width=2](12,0)--(14,0);
  \draw[-, line width=2](13,1)--(13,-1);

\fill[color=red!60, fill=red, fill opacity=1.0, shading=ball, ball color=red] (16,0) arc[start angle=-180,  end angle=0,x radius=2, y radius=0.6 , draw opacity=0, fill opacity=1.0]
arc[start angle=0, end angle=180, x radius=2, y radius=2,draw opacity=0, fill opacity=1.0];
;
\shade[color=blue!60,fill opacity=1.0, ] (16,0) arc[start angle=-180, end angle=0, x radius=2, y radius=2,draw opacity=0, fill opacity=1.0] arc[start angle=0,  end angle=180,x radius=2, y radius=-0.4 , draw opacity=0, fill opacity=1.0] ;
\fill[color=blue!60,fill opacity=1.0, ] (16,0) arc[start angle=-180, end angle=0, x radius=2, y radius=2,draw opacity=0, fill opacity=1.0] arc[start angle=0,  end angle=180,x radius=2, y radius=-0.4 , draw opacity=0, fill opacity=0.9] ;
 % \draw (22, 0) node[anchor=west , fontscale=3]\text[fontscale=3]{%Stokes' Law\\ 
 %    $+$};
     \draw[-, line width=2](22,0)--(24,0);
  \draw[-, line width=2](23,1)--(23,-1);
    \fill[fill=blue, fill opacity=1., shading=ball] (28,0) circle(2);
\fill[fill=red, fill opacity=1.0] 
(26,1.0) 
arc[start angle=-180, end angle=30, x radius=2, y radius=0.0, , draw opacity=0, fill opacity=0.0] 
arc[start angle=30, end angle=150, x radius=2, y radius=2, , draw opacity=0, fill opacity=1.0] 
arc[start angle=150, end angle=30, x radius=2, y radius=0.8, , draw opacity=0, fill opacity=1.0] 
;
 \fill[fill=red, fill opacity=1.0, ] 
(26,-1.0) 
arc[start angle=-180, end angle=30, x radius=2, y radius=0.0, , draw opacity=0, fill opacity=0.0] 
arc[start angle=-30, end angle=-150, x radius=2, y radius=2, , draw opacity=0, fill opacity=1.0] 
arc[start angle=-150, end angle=-30, x radius=2, y radius=0.8, , draw opacity=0, fill opacity=1.0] 
;
       \draw[-, line width=2](31,0)--(33,0);
  \draw[-, line width=2](32,1)--(32,-1);
  \draw[-, line width=2](34,-1)circle(0.1);
  \draw[-, line width=2](34.5,-1)circle(0.1);
  \draw[-, line width=2](35,-1)circle(0.1);
    \draw (37, 0) node(formula){$\Bigg)$};
\end{tikzpicture}

  \caption{Schematic of $T({\bf r})$, the temperature over a spherical particle surface and its multipole decomposition.}
 \label{fig:heat_moments_decomp} 
 % \label{fig:hotsp}
\end{figure}
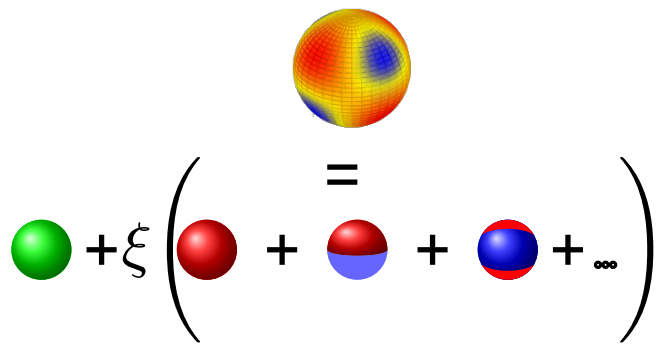

When a colloidal particle is heated to a small extent, there will be corrections to the force and torque acting on it. The local viscosity changes were described in~\cite{opp16} using heat moments, akin to a multipole expansion for charged particles in electrostatics \cite{jackson98}. An example is given in~\cref{fig:heat_moments_decomp}, which represents a schematic visualization of $T({\bf r})$, the temperature over the surface of a spherical particle and its multipole decomposition. In there, the force and torque were derived using  Lorentz reciprocal theorem \cite{Kim91}, an integral equation that relates the energy dissipation between different configurations~\cite{masoud19}. This procedure is commonly used to derive global-properties of viscous fluids, see, for example \cite{Stone1996,janus21,elfring_2017, nasouri2021minimum,Lauga2016}. For a fluid with a fixed viscosity gradient, this method was used to calculate the mobility tensor \cite{ziegler2022hydrodynamic}. This method has also been applied to the
calculation of the stresslet and the effective viscosity for slip-stick
particles~\cite{janus22} and for the
pressure drop~\cite{louis2023effect} under the influence of a temperature-dependent viscosity.

We extend the work presented in \cite{opp16} to find the stresslet of a hot particle, as well as the coupling between the torque and the shear rate of the fluid, enabling the calculation of the effective viscosity of a suspension of hot particles. We use both methods described above --- the energy dissipation and the stress-strain-rate relationship, which gives a full description of the suspension's rheology. The effective viscosity in such a case has two contributions --- one comes from the mean change of viscosity due to the average temperature increase coming from the heated particles. The second comes from the stresslets on the particles, which require modification due to the temperature field. 
 As we explain in what follows, the combined effect amounts to an effective viscosity given by
 \begin{equation}\label{eq:main_result}
{\eta^\mathrm{eff}}=
 \eta^{\infty}\left(1+2.5\varphi \right) 
 - 
 \xi \varphi \left(\eta^{\infty} \frac{V}{v^\mathrm{p}}\left\langle\mathcal{T}({\bf r})\right\rangle +
\frac{1}{v^\mathrm{p}}
 \frac{\boldsymbol{\mathcal{S}}^{ \xi}:{\bf E^{\infty}}}{\bf E^{\infty}:\bf E^{\infty}}  \right),
\end{equation}
where $\xi$ is a small parameter related to the change of viscosity with temperature, the brackets are defined as the average over the volume of the suspension, $V$, $\langle \mathcal{X}\rangle \equiv\frac{1}{V}\int\mathcal{X}dV$. 
The first term takes into account the effect of the local-temperature changes on the global viscosity, where $\mathcal{T}({\bf r})$ is the normalized-temperature (see below).
The second term is related to the correction to the stresslet, which is noted by $\boldsymbol{\mathcal{S}}^{ \xi}$. The particle's volume is $v^\mathrm{p}$, the double dot, $:$, is the second-order inner product ${\bf A}{:}{\bf B}=A_{ij}B_{ji}$, and throughout the work we use Einstein's notation for summation over repeated indices.
%Note that the first term in the correction is an extensive property.
For example, for a suspension of uniformly heated spherical particles, we show that by using Eq.~\ref{eq:main_result}, the effective viscosity is
\begin{equation}\label{eq:final_result_uniform}
    \frac{\eta^\mathrm{eff}}{\eta^{\infty}}=1+\frac{5}{2}\varphi-\xi \varphi\left(\frac{V}{v^\mathrm{p}}\langle \mathcal{T} \rangle+\frac{31}{144}\right).
\end{equation}
For a non-uniform heat distribution on the particles' surface, it is possible to express the correction to the viscosity
using an expansion of heat multiples in terms of spherical harmonics, as we show later on. In such a case, if the particles are Brownian and their orientation is random, all contributions from higher-order multipoles average out. If the particles are all oriented, there are additional contributions. The fluid is no longer isotropic and the viscosity is, in general, tensorial in nature. 

In what follows, we outline the derivation of Eq.~~\ref{eq:main_result} and its extensions, and give examples of its use. The remainder of the paper is structured as follows: Sec.~\ref{section:Fundamentals} presents fundamental concepts and defines the analytical framework for hot-particle analysis, followed by the determination of effective stress and viscosity for a suspension of hot particles in Sec.~\ref{section:eff_vis_hot_part}.
Then, we discuss these results and their consequences and give examples in Sec.~\ref{section:discussion}. 

\section{Background}\label{section:Fundamentals}

Let us first provide the foundational terminology necessary for describing a suspension of particles under conditions where inertial effects are negligible. Specifically, we consider a fluid undergoing steady, laminar shear. We then go on to outline Lorentz Reciprocal theorem for a hot suspension and the temperature and viscosity distribution around hot particles. We denote the velocity and pressure fields of the fluid by ${\bf u}({\bf r})$ and $p({\bf r})$ respectively. The stress tensor is given by
\begin{equation}
   { \boldsymbol \sigma} = -p{\bf I} +2\eta {\bf E},
\end{equation}
where $\eta$ is the fluid's viscosity and ${\bf E}$ is the symmetric-traceless shear-rate tensor defined as ${\bf E} =\frac{1}{2}\left( \boldsymbol \nabla {\bf u} + \left(\boldsymbol \nabla {\bf u}\right)^T\right)$. The flow obeys Stokes equations \cite{Kim91}:
\begin{eqnarray}
    &\boldsymbol \nabla \cdot {\bf u}=0, \label{eq:eq_incompressibility}\\
    & \boldsymbol \nabla \cdot \boldsymbol \sigma =0.\label{eq:eq_momentum}
\end{eqnarray}
Equation~\ref{eq:eq_incompressibility} comes from mass conservation for an incompressible fluid, and Eq.~\ref{eq:eq_momentum} comes from momentum conservation. The energy dissipation of a fluid in a volume $V$ surrounded by an area $S$ is defined as \cite{Bird60}
\begin{equation}
    \Phi=\int_{V}\boldsymbol \sigma:\boldsymbol \nabla { \bf u} dV =\int_{V}2\eta{\bf E}:{ \bf E}dV,
\end{equation}
where the second equality holds since ${\bf E}$ is a symmetric-traceless tensor. Utilizing the Divergence Theorem, we can rewrite the energy dissipation as
\begin{equation}
    \Phi = \int_{V}\boldsymbol \sigma:\boldsymbol \nabla {\bf u} \, dV = \int_{V} \boldsymbol \nabla \cdot \left(\boldsymbol \sigma \cdot { \bf u} \right) dV 
    = \int_{S}\boldsymbol \sigma\cdot{\bf n} \cdot {\bf u} \,dS,
\end{equation}
where in the second equality, we have used Eq.~\ref{eq:eq_momentum} and ${\bf n}$ is a unit vector normal to the surface.
Throughout this paper, we distinguish between the particulate fluid velocity ${\bf u}$, stress tensor $\boldsymbol \sigma$, and shear-rate tensor ${\bf E}$ and those of a bare fluid with no particles, for which we use the notation ${\bf u}^{\infty}$, $\boldsymbol\sigma^{\infty}$, ${\bf E}^{\infty}$.

\subsection{Lorentz Reciprocal Theorem}
Lorentz Reciprocal Theorem is a fundamental integral theorem for viscous fluids that connects two configurations --- a fluid with velocity $\boldsymbol u^\mathbf{1}$ and stress $\boldsymbol \sigma^\mathbf{1}$ and a fluid with velocity $\boldsymbol u^\mathbf{2}$ and stress $\boldsymbol \sigma^\mathbf{2}$,
\begin{equation}
\int_{S}\boldsymbol \sigma^\mathbf{1}\cdot {\bf n} \cdot {\bf u}^\mathbf{2} dS = \int_{S}\boldsymbol \sigma^\mathbf{2}\cdot { \bf n}\cdot   {\bf u}^\mathbf{1} dS,
    \label{eq:LorentzRecirocalOriginal}
\end{equation}
under the assumption that both fluids' velocities tend to zero on the volume's boundaries.
A possible interpretation of this theorem is that it connects the energy-dissipation of a stress field $\boldsymbol{\sigma}^\mathbf{1}$ acting on a fluid with velocity ${\bf u}^\mathbf{2}$ with respect to that of stress $\boldsymbol{\sigma}^\mathbf{2}$ and flow ${\bf u}^\mathbf{1}$, where $S$ and $V$ are the surface and volume being compared.

For two fluids of different viscosities, which can, in general, be spatially varying, where the first (second) fluid viscosity is $\eta^\mathrm{1}({\bf r)}$ [$\eta^\mathrm{2}({\bf r})$], there is an additional term that includes the viscosity difference and the shear-rate tensor $\bf E^\mathbf{1}$ ($ \bf E^\mathbf{2}$) of the first (second) fluid~\cite{opp16}
\begin{equation}
\int_{S}\boldsymbol \sigma^\mathbf{1}\cdot {\bf n} \cdot {\bf u}^\mathbf{2} dS= \int_{S}\boldsymbol \sigma^\mathbf{2}\cdot { \bf n}\cdot   {\bf u}^\mathbf{1} dS 
+ \int_{V} 2(\eta^\mathrm{1}({\bf r})-\eta^\mathrm{2}({\bf r})){\bf E}^\mathbf{1}:{\bf E}^\mathbf{2} dV.
\label{eq:LorentzRecirocal}
\end{equation}

\subsection{Temperature Distribution}

We describe the temperature field $T({\bf r})$ surrounding a spherical particle with radius $a$, exhibiting a deviation $\Delta T({\bf r})$ from the surrounding fluid, akin to \cite{opp16}. Assuming a small temperature difference, the viscosity can be expanded using a Taylor series. The particles act as sources of the heat flux vector ${\bf q}({\bf r})=-k_0\nabla T({\bf r})$, where $k_0$ denotes the thermal conductivity, and the heat-flux sources 
$\mathbf{\rho}^q({\bf r})=\nabla\cdot{\bf q}({\bf r})$ can be expanded into multipole terms, analogous to electrostatic expansion (see \cite{jackson98}). We note the zeroth, first, second and $m$-th order terms ($\Delta T^{{Q}^\mathrm{0}} = Q^0/(4\pi k_0 a)$, $\Delta T^{Q^\mathrm{1}}$, $\Delta T^{Q^\mathrm{2}}, T^{Q^\mathrm{m}}$, respectively) which represent the monopole, dipole, quadrupole and $m$-th order multipole heat moments given by
\begin{eqnarray}
&&Q^\mathrm{0}=\int \mathbf{\rho}^q({\bf r},\theta,\phi)dV, \\ 
&&\mathbf{Q}^\mathrm{1}=\frac{1}{\Delta T^{Q^\mathrm{1}}}\frac{1}{4\pi k_0 a ^2}\int {\bf r} \mathbf{\rho}^q({\bf r},\theta,\phi) dV \nonumber\\
&&\mathbf{Q}^\mathrm{2}=\frac{1}{\Delta T^{Q^\mathrm{2}}}\frac{1}{4\pi k_0a^3}\int \left({\bf r}{\bf r}-\frac{1}{3}{\bf I}\right)\mathbf{\rho}^q({\bf r},\theta,\phi)dV \nonumber,\\
&&\mathbf{Q}^\mathrm{m}=\frac{1}{\Delta T^{Q^\mathrm{m}}}\frac{1}{4\pi k_0a^m} \int \overbracket{{\bf r}_1...{\bf r}_m} \mathbf{\rho}^q({\bf r},\theta,\phi)dV \nonumber.
\end{eqnarray}
For higher orders we have used the notation \cite{brenner1964stokes, hess1984decay, hess2015, elfring2022active}, $\overbracket{{\bf r}_1...{\bf r}_l}=\frac{\left(-1\right)^l}{\left(2l-1\right)!!}\boldsymbol{\nabla}^n\left(\frac{1}{r}\right)_a$  which indicates the $n$-adic ``solid spherical
harmonic” of degree $-(l + 1),$ which is the \textit{irreducible} symmetric traceless
part of the $m$-th rank tensor. Then, considering energy conservation, neglecting the effects of radiation and convection and for a small P\'eclet number (such that distortions induced by the flow on the temperature field can be
neglected, and particle-scale features of the temperature distribution are imprinted on the thermal
field in the surrounding fluid), the diffusion heat transfer equation is  \cite{leal07}
\begin{equation}\label{eq:eqT}
\mathbf{\rho}^q({\bf r}) =-k_0 \nabla^2 T({\bf r}).
\end{equation}
This equation can be solved using spherical harmonic moments of the heat flux (see \cite{hess2015}). The solution to \cref{eq:eqT} can be expressed as
\begin{equation}\label{eq:eq_ts}
\begin{aligned} 
T({\bf r})&=
T^{\infty}+\sum_{m=1} \frac{1}{m!}\frac{1}{r^{m-1}}\overbracket{{\bf n}_1...{\bf n}_m}\mathbf{Q}^\mathrm{m}\\
&=T^{\infty}+\Delta T^{Q^\mathrm{0}} \frac{a}{r}+
\Delta T^{Q^\mathrm{1}} \frac{a^2\mathbf{Q}^\mathrm{1}\cdot {\bf n}}{ r^2}+
\Delta T^{ Q^\mathrm{2}} \frac{a^3\mathbf{Q}^\mathrm{2}:\left({\bf n}{\bf n}-\frac{1}{3} {\bf I}\right)}{2 r^3}+...
% \label{eqT}
\end{aligned}
\end{equation}

Assuming a small temperature difference, the viscosity around the particle can be expanded by a Taylor series (see also \cite{hess2015, brenner1964stokes, elfring2022active})
\begin{equation}\label{eq:pert_visc}
\begin{aligned}
        \eta^\mathrm{p}({\bf r})&=\eta^{\infty}+\sum_{m}\Delta T^{Q^\mathrm{m}}=\eta^{\infty}+\Delta T^{Q^\mathrm{0}} \left.\frac{\partial\eta}{\partial T}\right|_{T^{\infty}} \mathcal{T}({\bf r})\\
    &=\eta^{\infty}(1- \xi \mathcal{T}({\bf r})), 
\end{aligned}
\end{equation}
where $\xi \equiv -\Delta T^{Q^\mathrm{m}}\left.\frac{\partial \eta}{\partial T}\right|_{T^{\infty}}$ is assumed to be small (and positive for an increased temperature), and $T^{Q^\mathrm{m}}$ is defined by the leading order multipole (e.g. if there is a monopole it will be given by $T^{Q^\mathrm{0}}$). The dimensionless temperature distribution is defined as
\begin{equation}\label{eq:norm_temp}
    \mathcal{T}({\bf r})\equiv\frac{1}{\Delta T^{Q^m}}\left(T({\bf r})-T^{\infty}\right).
\end{equation}

\section{Effective Viscosity for a Suspension of Hot Particles}
\label{section:eff_vis_hot_part}
In this section, we calculate the corrections to Einstein's relation (Eq.~\ref{eq:Eins_rel}) coming from the temperature distribution over the particles' surface.
To compute this correction, we will use Eq.~\ref{eq:eff_vis_brenner}, i.e. 
$\eta^\mathrm{eff}=\eta^{\infty}\frac{\Phi}{\Phi^{\infty}}$.
The energy dissipation of a bare viscous homogeneous fluid, $\Phi^{\infty}$, 
in a container of volume $V$  is given by
% Let the fluid velocity and pressure $u_{free},p_{free}$. 
\begin{equation}
        \Phi^{\infty}=\int_V2\eta^{\infty} { \bf E}^{\infty}:{\bf E}^{\infty} dV=2\eta^\infty{\bf E}^{\infty}:{\bf E}^{\infty}V.
\end{equation}

The energy dissipation in the presence of hot particles, $\Phi$, is separated into two contributions --- on the boundary of the volume (``bulk fluid contribution") and around the particles themselves (``hot stresslet contribution") 
\begin{equation}
\Phi = \Phi^\mathrm{b}+\Phi^\mathrm{p}.    
\end{equation}
%as $\Phi_{\rm fluid\,with\,hot\,particles} = \Phi^\mathrm{p} + \Phi^\mathrm{b}.$
For both terms we use the extended Lorentz reciprocal theorem, Eq.~\ref{eq:LorentzRecirocal}, which takes into account the local-perturbation to the viscosity as represented in Eq.~\ref{eq:pert_visc}, where the normalized-temperature distribution $\mathcal{T}({\bf r)}$ is given by the heat multipole expansion as in \cref{eq:eq_ts}. 

Assuming the viscosity difference, $\xi$, is small, we will expand the shear-rate tensor in powers of $\xi$, using a perturbation expansion. Terms of order $O(\xi^2)$ will be neglected. The shear-rate tensor can be represented as a summation of two terms --- the free fluid shear-rate tensor plus the effect of the disturbance due to the presence of the particles, which includes the change to the viscosity. The disruption of the shear-rate tensor due to the particles can also be divided into two contributions --- the first, $\widetilde{\bf E}^{(0)}$, represents the shear-rate tensor disruption due to the presence of a ``cold" particle. The second, $\widetilde{\bf E}^{(1)}$, represents the correction to the shear-rate tensor that is related to local viscosity changes. This second term is of order $\xi$~\cite{opp16}
\begin{equation}\label{eq:shear_to_dist}
\begin{aligned}
             {\bf E}({\bf r})&=  {\bf E}^{\infty} + \widetilde{\bf E}({\bf r})\\
         &={\bf E}^{\infty}+ \widetilde{\bf E}^{(0)}({\bf r})+\xi \widetilde{\bf E}^{(1)}({\bf r}).
\end{aligned}
\end{equation}
The total shear rate tensor, ${\bf E}$, must be proportional to the shear rate tensor of the bare fluid, ${\bf E}^{\infty}$. We thus define the normalized  4-th rank 
 shear-rate tensor $\boldsymbol{\mathcal{E}}$ as
\begin{equation}\label{eq:norm_str_def}
    { E}_{ij}({\bf r})={\mathcal{E}}_{ijkl}({\bf r}){ E}^{\infty}_{kl}.
\end{equation}
In particular, we can represent the different terms in \cref{eq:shear_to_dist} using the normalized strain-rate tensor
$\widetilde{\bf E}(\boldsymbol r)=\widetilde{\boldsymbol{\mathcal{E}}}(\boldsymbol r):{\bf E}^{\infty}$.
In the next subsections, we derive separately each of the terms in $\widetilde{\bf E}$. First, the bulk contribution in Sec.~\ref{subsection:boundary_cont}, then the contribution of the particles in Sec.~ \ref{subsection:particles_cont}.

\subsection{Bulk Fluid Contribution}\label{subsection:boundary_cont}
%============================================
In considering the bulk, we examine the energy dissipation, represented as
\begin{equation}\label{eq:phi_boundary}
    \Phi^\mathrm{b} = \int_{S}\boldsymbol \sigma \cdot {\bf n} \cdot  {\bf u}^{\infty} \,dS,
\end{equation}
where ${\bf u}^{\infty}$ denotes the velocity on the boundary, $S$, that has not changed due to viscosity variations. 
We would like to relate $\Phi^\mathrm{b}$ to the dissipation of a particulate-free fluid $\Phi^{\infty}$. To account for viscosity differences due to the presence of hot particles, we employ Lorentz's reciprocal theorem (Eq.~\ref{eq:LorentzRecirocal}). We express $\Phi^\mathrm{b}$ as
\begin{equation}
\begin{aligned}
    \Phi^\mathrm{b} &=\int_{S}\boldsymbol \sigma^{\infty} \cdot {\bf n} \cdot {\bf u}  dS 
    +\int_{V}2 \Delta \eta  {\bf E}^{\infty}:{ \bf E} dV \\
     &= \Phi^{\infty} + \Phi^{\Delta \eta},
\end{aligned}    
\end{equation}
where $\Delta \eta({\bf r}) = \eta({\bf r}) - \eta^{\infty}$, and the second term is labeled $\Phi^{\Delta \eta}$. 
We proceed to relate $\Phi^{\Delta \eta}$ to $\Phi^{\infty}$. The local viscosity of the bulk fluid is changed due to the presence of hot particles. It can be written as a sum of all the temperature differences on all particles. From Eq.~\ref{eq:pert_visc} we have
\begin{equation}
\begin{aligned}
    \eta({\bf r}) &= \eta^{\infty}\left(1-\sum_{\mathbf{p}} \xi \mathcal{T}({\bf r}-{\bf r}^\mathbf{p})\right ).\label{eq:bulk_local_visc}
\end{aligned}
\end{equation}

For a dilute homogeneous suspension% of spherical particles
, neglecting mutual effects between particles and the correction to the shear rate coming from viscosity changes, ${\bf E}^{(1)}$, which gives a correction of $O(\xi^2)$ to the energy dissipation, the shear-rate contribution from particles is given by
\begin{equation}
\label{eq:E_with_particles}
\begin{aligned}
    { E}_{ij}&=\left(\delta_{ik}\delta_{jl}+\sum_{p}\mathcal{E}^\mathrm{p}_{ijkl}({\bf r}-{\bf r}^\mathbf{p})\right){ E}_{kl}^{\infty},
\end{aligned}
\end{equation}
where ${\boldsymbol{\mathcal{E}}^\mathbf{p}}({\bf r})$ is the normalized strain-rate tensor, as defined in Eq.~\ref{eq:norm_str_def}, for each particle. 
Using Eq.~\ref{eq:E_with_particles} we can express $\Phi^{\Delta \eta}$ as

\begin{equation}
\begin{aligned}
    \Phi^{\Delta \eta} =&-\int_V 2 \eta^\infty { E}_{ij}^{\infty}{ E}_{kl}^\infty\xi \sum_{\mathbf{p}} \mathcal{T}({\bf r}-{\bf r}^\mathbf{p})\left(\delta_{ik}\delta_{jl}+\sum_{\mathbf{p}_1}{\mathcal{E}}^{p_1}_{ijkl}({\bf r}-{\bf r}^{\mathbf{p}_1})\right)dV
    \\=& -\frac{\Phi^{\infty}}{V}\xi \sum_{\mathbf{p}} \int_{V} \mathcal{T}({{\bf r}-{\bf r}^\mathbf{p}})dV-\xi{ E}_{ij}^{\infty}{ E}_{kl}^{\infty}
\sum_{\mathbf{p}}\sum_{\mathbf{p}_1}\int_{V}
 \mathcal{T}({{\bf r}-{\bf r}^\mathbf{p}}){{\mathcal{E}_{ijkl}^{p_  1}}}({\bf r}-{\bf r}^{\mathbf{p}_1})dV.
\end{aligned}
\end{equation}
Throughout this work, we will assume a continuous density such that summation over the number of particles will be replaced by integration over the volume $dV^{p}$ with
\begin{equation}\label{eq:continoum_assumption}
   dv^{p} = \varphi ({\bf r^p}) dV.
   \end{equation}
The \textit{local} volume fraction 
$\varphi({\bf r}^\mathbf{p})$ is assumed constant and is given by
\begin{equation}
\label{eq:volume_fraction}
\varphi=\frac{N v^\mathrm{p}}{V},
\end{equation} 
where $N$ is the number of particles in the volume $V$, normalized by the particle's volume $v^\mathrm{p}.$
We can now see that the second term is negligible as it is of order $\varphi^2$. We are left with 
\begin{equation}
\label{eq:bulkDissipation}
\frac{{\Phi}^\mathrm{b}}{\Phi^{\infty}} =
  1-\xi{\varphi} \frac{V}{v^\mathrm{p}}\big\langle \mathcal{T}({\bf r})\big\rangle.
\end{equation}
The hot particles heat the fluid causing the viscosity to decrease. Notice that this term gives an extensive correction. For a given concentration, the larger the system, the hotter it will be. 
\subsection{Hot Particle Contribution}\label{subsection:particles_cont}
We will now compute the contribution to the effective viscosity coming from the stresslet and the torque, under the assumption of a dilute suspension such that the energy dissipation from each particle is independent of its neighbors and can be calculated separately,
     \begin{equation}
              \Phi^\mathrm{p} = \mathlarger{\sum}_{\rm p} \int_{S^\mathbf{p}}\boldsymbol\sigma\cdot{\bf n}\cdot{\bf u}^\mathbf{p}dS. 
     \end{equation}
For its calculation, we first model the motion of a particle in the fluid, assuming it translates with velocity $\bf U$ and rotates with angular velocity $\boldsymbol{\Omega}=\frac{1}{2}\boldsymbol{\varepsilon}\cdot\boldsymbol{\omega}$, where $\boldsymbol{\varepsilon}$ is the permutation symbol. We also assume linear-flow of the fluid in the absence of particles \cite{Kim91}
\begin{equation}
\label{eq:linear_flow}
{\bf u}^{\infty}={\bf U}^{\infty}+\boldsymbol{\Omega}^{\infty}\times{\bf r}+{\bf E}^{\infty}\cdot{\bf r}.
\end{equation}
%then each particle velocity 
The disturbance flow calculated on the particle's boundary is given by
\begin{equation}
    \left.{\bf u}^\mathbf{p}\right|_{S^\mathbf{p}} = ({\bf U}^{\infty}-{\bf U})+({\bf \Omega}^{\infty}-\boldsymbol{\Omega})\times{\bf r}+{\bf E}^{\infty}\cdot {\bf r}.
\end{equation} 
The contribution of the disturbance flow to the effective viscosity is given by
\begin{equation}
     \int_{S^\mathbf{p}}\boldsymbol\sigma\cdot{\bf n}\cdot{\bf u}^\mathbf{p}dS=
    {\bf F}\cdot({\bf U}^{\infty}-{\bf U})+
    {\bf T}\cdot({\bf \Omega}^{\infty}-\boldsymbol{\Omega})+
    \boldsymbol{\mathcal{S}}: {\bf E}^{\infty},\label{eq:ene_dispp_part}
\end{equation}
where $\bf F$ is the total force
${\bf F}=\int_{S^p} \boldsymbol{\sigma} \cdot {\bf n} dS$, the torque is ${\bf T} =\int_{S^\mathbf{p}}{\bf r} \times\left(\boldsymbol{ \sigma} \cdot {\bf n}\right)  dS$,  and the symmetric-traceless part of the force-dipole is the \textit{stresslet}, 
\begin{equation}
        \mathcal{S}_{ij}=\int_{S^p} \left[\frac{1}{2}\left(\sigma_{ik} n_k  r_j+\sigma_{jk} n_k  r_i \right) - \frac{1}{3}\sigma_{lk} n_k  r_l\delta_{ij}  \right]dS.
        \end{equation}

In order to find the correction to the stresslet and the torque due to the temperature field created by the particles, we follow \cite{Lauga2016,opp16} in using an extension of Lorentz reciprocal theorem, \cref{eq:LorentzRecirocalOriginal}, for computing the force moments of a particle in a viscous fluid (see also \cite{Stone1996,elfring_2017, janus21, janus22, masoud19}). 
We employ an auxiliary problem of a ``cold particle" (marked with superscript $X^{\rm cold}$), with the same surface temperature as its environment and a linear movement profile. We compare it to our desired problem (which we mark with superscript $X^{\rm hot}$). The  \textit{disturbance} fluid velocity on the particles' boundaries is given by
\begin{eqnarray}
&&{\bf u}^{\rm hot}={\bf U}^{\rm hot}+\boldsymbol{\Omega}^{\rm hot}\times{\bf r}+{\bf E}^{\rm hot}\cdot{\bf r} \nonumber  \\
&&{\bf u}^{\rm cold}={\bf U}^{\rm cold}+\boldsymbol{\Omega}^{\rm cold}\times{\bf r}+{\bf E}^{\rm cold}\cdot{\bf r}.
\end{eqnarray}
The disturbance velocities obey Lorentz reciprocal theorem of the form of Eq.~\ref{eq:LorentzRecirocal}, where the local viscosity is given by \cref{eq:pert_visc}, which gives,
\begin{equation}
\label{eq:LorentzColdHot}
    {\bf U}^{\rm cold}\cdot {\bf F}^{\rm hot}+
    \boldsymbol{\Omega}^{\rm cold}\cdot{\bf T}^{\rm hot} +{\bf E}^{\rm cold}: \boldsymbol{\mathcal{S}}^{\rm hot}=
    {\bf F}^{\rm cold}\cdot{\bf U}^{\rm hot} +\boldsymbol{\Omega}^{\rm hot}\cdot{\bf T}^{\rm cold}+\boldsymbol{\mathcal{S}}^{\rm cold}: {\bf E}^{\rm hot}-
    2\eta^{\infty}\xi\int  \mathcal{T}({\bf r}){\bf E}^{\rm cold}:{\bf E}^{\rm hot} dV.
\end{equation}
Without loss of generality, we substitute in Eq.~\ref{eq:LorentzColdHot} the force-velocity, torque-angular-velocity, and stresslet-shear-rate relations for a cold-particle, where there is no coupling between different velocity gradients and force moments,
\begin{eqnarray}
\label{eq:forces_velocities}
    {\bf F}^{\rm cold}= {\bf U}^{\rm cold}\cdot{\boldsymbol{\mathcal{R}}}^{{{\bf F}{\bf U}},{\rm cold}}\\
{\bf T}^{\rm cold}=\boldsymbol{\Omega}^{\rm cold}\cdot{\boldsymbol{\mathcal{R}}}^{{{\bf T}\boldsymbol\Omega},{\rm cold}} \nonumber \\ 
{{\boldsymbol {\mathcal{S}}}}^{\rm cold}={\bf E}^{\rm cold}:{\boldsymbol{\mathcal{R}}}^{{ \boldsymbol {\mathcal{S}}{\bf E}},{\rm cold}}. \nonumber \end{eqnarray}
We can write Eqs.~\ref{eq:forces_velocities} compactly using the grand-resistance tensor \cite{Kim91}. For example, for a spherical rigid particle 
\begin{equation}
\label{eq:resistance_cold}
    \begin{bmatrix}
    F^{\rm cold}_i\\T_i^{\rm cold}\\S_{ij}^{\rm cold}
    \end{bmatrix} = \eta^{\infty}
    \begin{bmatrix}
    6\pi a \delta_{ij}& 0& 0 \\
     0 &8\pi a^3 \delta_{ij}& 0 \\
     0 & 0& \frac{20 \pi a^3}{3}\delta_{ik}\delta_{jl}
    \end{bmatrix}
    \begin{bmatrix}
    U_j^{\rm cold}\\
    \Omega_j^{\rm cold}\\
    E_{kl}^{\rm cold}
    \end{bmatrix}.
\end{equation}

We will now use the normalized strain-rate tensor as represented in \cref{eq:norm_str_def} as well as its equivalents for the force and torque 
\begin{equation}
E_{nm}({\bf r})= {\mathcal{E}}_{inm}^{{\rm cold},U}({\bf r})U_i \ \ \ ; \ \ \ 
E_{nm}({\bf r})= 
{\mathcal{E}}_{inm}^{{\rm cold},\Omega}({\bf r})\Omega_i. 
\end{equation}
We will also use the linear-separability of Eq.~\ref{eq:LorentzColdHot} with respect to $
{\bf U}^{\rm cold},\boldsymbol{\Omega}^{\rm cold}$ and ${\bf E}^{\rm cold}$ which allows us to extract the grand resistance tensor
\begin{equation}
\label{eq:grandResistance}
    \begin{bmatrix}
    {\bf F}^{\rm hot}\\{\bf T}^{\rm hot}\\{{\boldsymbol {\mathcal{S}}}}^{\rm hot}
    \end{bmatrix}=
    {
    \begin{bmatrix}
    {\boldsymbol{\mathcal{R}}}^{{\bf F}{\bf U}}&{\boldsymbol{\mathcal{R}}}^{{\bf F}\boldsymbol\Omega}&{\boldsymbol{\mathcal{R}}}^{{\bf F}{\bf E}}\\
    {\boldsymbol{\mathcal{R}}}^{{\bf T}{\bf U}}&{\boldsymbol{\mathcal{R}}}^{{\bf T}\boldsymbol \Omega}&{\boldsymbol{\mathcal{R}}}^{{\bf T}{\bf E}}\\
    {\boldsymbol{\mathcal{R}}}^{ \boldsymbol {\mathcal{S}}{\bf U}}&
    {\boldsymbol{\mathcal{R}}}^{ \boldsymbol {\mathcal{S}}\boldsymbol\Omega}&
    {\boldsymbol{\mathcal{R}}}^{ \boldsymbol {\mathcal{S}}{\bf E}}
    \end{bmatrix}}
    % _{\boldsymbol{\mathcal{R}}}
    \begin{bmatrix}
    {\bf U}^{\rm hot}\\
    \boldsymbol{ \Omega}^{\rm hot}\\
    {\bf E}^{\rm hot}
    \end{bmatrix},
\end{equation}
then for the stresslet we can separate the cold-particle contribution  ${{\boldsymbol {\mathcal{S}}}}^{\rm cold}$
and the perturbed stresslet coming from the local viscosity changes, as given by \cref{eq:LorentzColdHot}. We denote the perturbation to the stresslet $\boldsymbol{\mathcal{S}}^{ \xi}$. Similarly for the torque, the cold-particle contribution is ${\bf T}^{\rm cold}$ and the perturbed torque is marked ${\bf T}^\xi$, such that
\begin{eqnarray}
    \label{eq:stresslet_total}
    {{\boldsymbol {\mathcal{S}}}}^{\rm hot}={{\boldsymbol {\mathcal{S}}}}^{\rm cold}-\xi\boldsymbol{\mathcal{S}}^{ \xi},
\\
    \label{eq:torque_total}
    {\bf T}^{\rm hot}={\bf T}^{\rm cold}-\xi{\bf T}^{ \xi}.
\end{eqnarray}
For a pure shear-rate in the absence of forces and torques, ${\bf u}^{\rm hot}={\bf E}^{\rm hot}\cdot{\bf r}$ and the stresslet can be written as with the different terms driven from Eq.~\ref{eq:norm_str_def}, Eq.~\ref{eq:forces_velocities}, Eq.~\ref{eq:LorentzColdHot} and Eq.~\ref{eq:grandResistance}, such that
\begin{equation}\label{eq:stresslet_decom}
    {{\boldsymbol {\mathcal{S}}}}^{\rm hot}
    =\left({\boldsymbol{\mathcal{R}}^{{{\boldsymbol{\mathcal{S}}\bf E}},{\rm cold}}}-\xi\widetilde{\boldsymbol{\mathcal{R}}}^{\boldsymbol{\mathcal{S}}\bf E}\right):{\bf E}^{\rm hot}, 
\end{equation}
and the perturbed resistance tensor is given by
\begin{equation} \widetilde{\mathcal{R}}_{ijkl}^{\mathrm{SE}}=2\eta^{\infty}\int_Q \mathcal{T}({\bf r})
    {\mathcal{E}}_{ijnm}^{\rm cold}({\bf r}){\mathcal{E}}_{klnm}^{\rm cold}({\bf r}) dV.
    \label{eq:resistance_se}
\end{equation}
%, $\widetilde{\boldsymbol{\mathcal{R}}}^{\boldsymbol{\mathcal{S}}\bf E}$. 

With the model problem at hand, we can now extract the correction to the stresslet in \cref{eq:ene_dispp_part}. In this case $\boldsymbol{\mathcal{S}}^{\rm hot} =\boldsymbol{\mathcal{S}}$ and $\mathbf{E}^{\rm hot} = \mathbf{E}^\infty$. In the absence of external forces and torques, the stresslet is therefore given by Eq.~\ref{eq:stresslet_total} with the perturbation to the stresslet given by
\begin{equation}\label{eq:stresslet_correctness}
\mathcal{S}^{ \xi}_{ij}=\widetilde{\mathcal{R}}^{\mathcal{S}{\rm E}}_{ijkl}E^{\infty}_{kl}
\end{equation} 
The result is given by
(\cref{eq:stresslet_correctness}) is generic for non-slip rigid particles. In the next section, we present detailed examples of a spherical particle under different heat distributions. 
If the particle is not uniformly heated, there may be an additional contribution to the dissipation, which comes from an induced angular velocity coming from the shear rate itself, which introduces coupling between the angular velocity and the stresslet, even in the absence of forces and torques. Such coupling does not happen for cold spherical particles. It may appear for hot particles, but such a term is of order $O(\xi)$, and its net contribution to the stresslet is $O(\xi^2)$ and thus negligible.

Assuming a homogeneous suspension, we can use the volume fraction, \cref{eq:volume_fraction}, to get %in \cref{eq:phi_part} to write
\begin{equation}
%\begin{aligned} 
    \Phi^\mathrm{p} = 
    \varphi\frac{V}{v^\mathrm{p}}\left(\boldsymbol{\mathcal{S}}^{\rm cold}-\xi\,\boldsymbol{\mathcal{S}}^{ \xi}\right):{\bf E}^{\infty}.
%\end{aligned}
\end{equation}
Lastly, we can sum the two contributions, from the bulk fluid and from the stresslet, to calculate the total energy dissipation. The ratio of the total energy dissipation,$\Phi = \Phi^{\rm b} + \Phi^{\rm p}$, to the bare fluid energy dissipation, $\Phi^{\infty}$, gives the effective viscosity presented in \cref{eq:main_result}.
The effective viscosity given in Eq.~\ref{eq:main_result} is general and applies to any hot rigid particle, not necessarily spherical. In the next section, we give a few specific examples of spherical particles of various heat distributions. 

Let us also consider a case where the particles all have the same orientation. If the particles are oriented due to an external field such that they cannot rotate, there will be a torque acting on them, enforcing ${\bf \Omega} = 0$. We can find the torque from \cref{eq:grandResistance}
\begin{equation}\label{eq:hot_part_torque}
    {\bf T}=\left({\boldsymbol{\mathcal{R}}}^{{\bf T}\boldsymbol \Omega}-\xi\widetilde{{\boldsymbol{\mathcal{R}}}}^{{\bf T}\boldsymbol \Omega}\right)\cdot{\boldsymbol{\Omega}}^\infty-\xi\widetilde{\boldsymbol{\mathcal{R}}}^{{\bf T} {\bf E}}:{\bf E}^\infty,
\end{equation}
with
\begin{equation}\begin{aligned}
    &\widetilde{\mathcal{R}}_{ij}^{\mathrm{ T\Omega}}=2\eta^{\infty}\int_Q \mathcal{T}({\bf r})
    {\mathcal{E}}_{inm}^{{\rm cold},\Omega}({\bf r}){\mathcal{E}}_{jnm}^{{\rm cold},\Omega}({\bf r}) dV,
    \\
    &\widetilde{\mathcal{R}}_{ijk}^{\mathrm{ TE}}=2\eta^{\infty}\int_Q \mathcal{T}({\bf r})
    {\mathcal{E}}_{inm}^{{\rm cold},\Omega}({\bf r}){\mathcal{E}}_{jknm}^{{\rm cold}}({\bf r}) dV.
    \end{aligned}
\end{equation}
In such a case, there will be a correction to the dissipation, and thus to the viscosity, coming from the torque.
In addition, \cref{eq:stresslet_decom} should include an extra term related to the angular-velocity $\boldsymbol{\Omega}^\infty$
\begin{equation}
        {{\boldsymbol {\mathcal{S}}}}\label{eq:hot_part_stresslet}
    =\left({\boldsymbol{\mathcal{R}}^{{{\boldsymbol{\mathcal{S}}\bf E}}}}-\xi\widetilde{\boldsymbol{\mathcal{R}}}^{\boldsymbol{\mathcal{S}}\bf E}\right):{\bf E}^\infty-
    \xi\widetilde{\boldsymbol{\mathcal{R}}}^{\boldsymbol{\mathcal{S}}\boldsymbol{\Omega}}\cdot\boldsymbol{\Omega}^\infty
\end{equation}
where
\begin{equation}\widetilde{{\mathcal{R}}}^{{\mathcal{S}}{\Omega}}_{ijk}=2\eta^{\infty}\int_Q \mathcal{T}({\bf r})
    {\mathcal{E}}_{ijnm}^{{\rm cold}}({\bf r}){\mathcal{E}}_{knm}^{{\rm cold},\Omega}({\bf r}) dV.
\end{equation}
 Due to \cref{eq:LorentzRecirocal}, we expect symmetries between the different resistance tensors, such that \cite{hinch1972note, opp16, masoud19} 
\begin{equation}
    \widetilde{{\mathcal{R}}}^{{\mathcal{S}}{\Omega}}_{ijk}=\widetilde{\mathcal{R}}_{kij}^{\mathrm{ TE}}.
\end{equation}
Then, following \cite{brenner1970rheology}, the effective viscosity will have additional terms
\begin{equation}
\frac{\eta^{\rm eff}}{\eta^{\infty}} = \frac{\Phi}{\Phi^{\infty}} =   1 - \xi {\varphi}\frac{V}{v^\mathrm{p}}\big\langle \mathcal{T}({\bf r})\big\rangle+\frac{\varphi}{v^\mathrm{p}}\frac{\left(\boldsymbol{\mathcal{S}}^{\rm cold}-\xi\,\boldsymbol{\mathcal{S}}^{ \xi}\right):{\bf E}^{\infty}+\left({\bf T^{\rm cold}}-\xi {\bf T}^\xi\right)\cdot\boldsymbol{\Omega}^\infty}{2\left|{\bf E}^{\infty}\right|^2}.
\end{equation} 

\subsection{Tensorial Form of the Effective Viscosity}
Using energy dissipation allows us to calculate a scalar correction to the viscosity, but the full Rheological response of the fluid is missing. Batchelor \cite{batchelor1970stress} calculated the effective viscosity as the ratio of applied strain-rate to the bulk stress. This method gives similar results when the fluid is isotropic, but  allows us to easily extract a tensorial viscosity in cases where the fluid is no longer isotropic, such as when the particles are all oriented in the same direction. In his work, Batchelor defined the tensorial effective viscosity as
\begin{equation}\label{eq:eff_visc_tensor}\
     \langle{\Sigma}\rangle_{ij}=-p^\mathrm{eff}\delta_{ij}+{\eta}^\mathrm{eff}_{ijkl}\frac{\partial u^\infty_k}{\partial r_l}.
 \end{equation}
He showed that the effective viscosity has two contributions --- one coming from the stresslet and the other from the torque, which can be written as
\begin{equation}
\label{eq:averageStress}
\begin{aligned}
    \langle \boldsymbol{\Sigma}\rangle&=-p^\mathrm{eff}{\bf I}+2\eta^{\infty}{\bf E}^{\infty}+\frac{\varphi}{v^\mathrm{p}}\left(\langle \boldsymbol{\mathcal{S}}\rangle+\frac{1}{2}\boldsymbol{\varepsilon}\cdot\langle\mathbf{T}\rangle\right).
\end{aligned}
\end{equation}
The ratio between average stress and strain rate reproduces Einstein's formula (\cref{eq:Eins_rel}) for a cold colloidal suspension, in which case $\eta^\mathrm{eff}$ is scalar. 
For a suspension of hot particles, there are two corrections to Eq.~\ref{eq:averageStress}. We can write, 
\begin{equation}
    \langle \boldsymbol{\Sigma}\rangle=-p^\mathrm{eff}{\bf I}+\boldsymbol{\Sigma}^\mathrm{b}+\varphi \boldsymbol{\Sigma}^\mathrm{p}.
\end{equation}
The first term, $\boldsymbol{\Sigma}^\mathrm{b}$, comes from the bulk fluid being heated
\begin{equation}
\boldsymbol{\Sigma}^\mathrm{b}=\langle\eta {\bf E}^\infty\rangle=2\eta^{\infty}\left(1-\xi N\langle \mathcal{T}({\bf r})\rangle\right){\bf E}^{\infty}
 = 2\eta^{\infty}\left(1-\xi \varphi\frac{V}{v^p}\langle \mathcal{T}({\bf r})\rangle\right){\bf E}^{\infty}.
\end{equation}
The second term, $\boldsymbol{\Sigma}^\mathrm{p}$, includes the particle contribution to the bulk stress, which, using \cref{eq:hot_part_stresslet,eq:hot_part_torque}, is given by
\begin{equation}\label{eq:paticles_bulk_stress}
        \boldsymbol{\Sigma}^\mathrm{p}=\frac{1}{v^\mathrm{p}}\left[\left(\boldsymbol{\mathcal{R}}^{\boldsymbol{\mathcal{S}}\bf E}
-\xi\widetilde{\boldsymbol{\mathcal{R}}}^{\boldsymbol{\mathcal{S}}\bf E} 
    \right):{\bf E}^{\infty}-\xi\widetilde{\boldsymbol{\mathcal{R}}}^{\boldsymbol{\mathcal{S}}\boldsymbol\Omega}\cdot\boldsymbol{\Omega}^\infty+\right.\left.\frac{1}{2}\boldsymbol{\varepsilon}\cdot\left({\boldsymbol{\mathcal{R}}}^{{\boldsymbol{T}\boldsymbol{\Omega}}}-\xi\widetilde{\boldsymbol{\mathcal{R}}}^{{\boldsymbol{T}\boldsymbol{\Omega}}}\right)\cdot\boldsymbol{\Omega}^\infty-\frac{1}{2}\xi\boldsymbol{\varepsilon}\cdot\widetilde{\boldsymbol{\mathcal{R}}}^{\boldsymbol{T{\bf E}}}:{\bf E}^\infty\right].
\end{equation}
Then, following past works \cite{batchelor1970stress, brenner1970rheology, ramachandran2009dynamics, janus22}, the effective viscosity comes from an induced  simple shear flow, ${\bf u}^{\infty} = y \hat{x}$,  which includes a shear-rate tensor and a vorticity vector
\begin{equation}
\label{eq:simple_shear}
\begin{aligned}
        {\bf E}^{\infty}=\frac{\dot\gamma}{2}\begin{bmatrix}
        0&1&0\\1&0&0\\0&0&0
    \end{bmatrix}
    &\quad ; \quad \boldsymbol{\Omega}^{\infty}=\frac{\dot\gamma}{2}\begin{bmatrix}
        0\\0\\1
    \end{bmatrix}.
\end{aligned}
\end{equation}
The scalar effective viscosity is defined by the ratio
 \begin{equation}
 \label{eq:hisotric_viscsoity}
\eta^{\mathrm{p}}=\Sigma_{12}/\dot\gamma.
 \end{equation}
 As expected, \cref{eq:main_result} is also recovered by the Batchelor method.
We can also consider a tensorial effective viscosity as given by Eq.~\ref{eq:eff_visc_tensor}. In what follows, we do not explicitly write the full tensorial viscosity tensor as it is a fourth-order tensor with 81 terms. However, we do express the anisotropic nature of the suspension by examining the different terms of the effective stress tensor. For an isotropic suspension, the stress is simply proportional to the shear-rate by a scalar multiplier, which is not the case for a hot, anisotropic suspension. In the next section, we will give results for a few particular examples. We will present the effective viscosity as historically defined by Eq.~\ref{eq:hisotric_viscsoity}, but we will also examine the anisotropic nature by showing different terms in the stress tensor that do not appear in the shear-rate tensor. We will also showcase how applying a different simple shear will result in a different effective scalar viscosity, again highlighting the anisotropic nature. 
 %In what follows we will take $\dot{\gamma} = 1$ for simplicity.

\section{Consequences and examples}
\label{section:discussion}
Next, we will look at particular heat distributions and their effect on the viscosity.  
The viscosity has two contributions from the heated bulk fluid, $\eta^{\rm b} = \Phi^{\rm b}/\Phi^{\infty}$ and the stresslet, $\eta^{\mathrm{p}}=\Phi^{\mathrm{ p}}/\Phi^{\infty}$, such that combined
\begin{equation}
    % \eta^{\rm eff}_{ijkl} = \eta^{\mathrm{b}}\delta_{ik}\delta_{jl} + \eta^{\mathrm{ p}}_{ijkl}.
    \eta^{\rm eff} = \eta^{\mathrm{b}} + \eta^{\mathrm{ p}}.
\end{equation}
The first part, coming from the bulk fluid, is always given by \cref{eq:bulkDissipation}. We present below the correction term coming from the stresslet. 
To apply our analysis, we focus on the effective viscosity of a suspension of spherical particles. We present a few examples and their consequences. 
First, to calculate the resistance tensor given in \cref{eq:resistance_se} for a spherical particle, we can use the normalized strain-rate tensor for a rigid cold sphere \cite{Datt19}
\begin{equation}
\label{eq:norm_shear_sphere}
\begin{aligned}
\mathcal{E}^{{\rm cold}}_{ijkl}({\bf r}) &= 
\frac{a^5}{2r^5}\left(\delta_{ik}\delta_{jl}+\delta_{jk}\delta_{il}\right)+
\frac{5a^3}{2r^5}\delta_{kl}r_ir_j
+\frac{5a^3}{4r^5}\left(\delta_{il}r_jr_k + \delta_{ik}r_jr_l + \delta_{jl}r_ir_k + \delta_{kj}r_ir_l\right) \\&+
\frac{5a^5}{2r^7}\left(\delta_{kl}r_ir_j + \delta_{jl}r_ir_k +\delta_{il}r_jr_k + \delta_{jk}r_ir_l + \delta_{ik}r_jr_l\right)
+5\left(\frac{7a^5}{2r^9}-\frac{5a^3}{2r^7}\right)r_ir_jr_kr_l.     
     \end{aligned}
     \end{equation}
The normalized strain-rate tensor for the angular-velocity is \cite{opp16}
     \begin{equation}
        {\mathcal{E}}_{inm}^{{\rm cold},\Omega}({\bf r})= -\frac{3a^3}{2r^5} (r_nr_j\varepsilon_{mji} + r_mr_j \varepsilon_{nji}).
     \end{equation}
% A substitution of the normalized strain-rate tensor for a spherical particle (Eq.~\ref{eq:norm_shear_sphere}), not only the monople, dipole and qudropole are required, but also the forth-order moment, 
Next, \cref{eq:resistance_se} requires the temperature distribution on the particle, $\mathcal{T}$, which is given by Eq.~\ref{eq:eq_ts} and \cref{eq:norm_temp}. Most generally, not only the heat monopole ($Q^\mathrm{0}$) is required, but also quadrupole ($Q^\mathrm{2}$) and the hexadecapole (which is marked by~$Q^\mathrm{4}$, giving a characteristic change in temperature $\Delta{T}^{Q^\mathrm{4}}$). As it happens, the dipole ($Q^\mathrm{1}$) and the octopole ($Q^\mathrm{3}$) do not contribute to the effective viscosity as well as all higher orders of the temperature distribution of the sphere.
Using Wolfram-Mathematica \cite{Mathematica}, and specifically the package given in ~\cite{einarsson2017computer}, we can calculate the integration in \cref{eq:resistance_se}, giving the resistance tensor for a hot sphere $\widetilde{\boldsymbol{\mathcal{R}}}^{\boldsymbol{\mathcal{S}}{\bf E}}$ 
\begin{align}\label{eq:resistance_stresslet}
\widetilde{{ \mathcal{R}}}^{{\boldsymbol{\mathcal{S}}\bf E}}_{ijkl}=&\frac{4\eta ^{\infty} \pi  a^3}{3}\Big[
\frac{19   \delta _{il} \delta _{jk}}{16 }-
\frac{13   \delta _{ik} \delta _{jl}}{24 }+
\frac{19   \delta _{ij} \delta _{kl}}{16 }+\frac{1}{28}\frac{ {\Delta T}^{Q^\mathrm{2}} }{{\Delta T}^{Q^\mathrm{0}}}
\left( {\delta _{ij} Q^{\mathrm{2}}_{kl}}-
\frac{61}{30}\delta _{ik} Q^{\mathrm{2}}_{jl}+
{\delta _{il} Q^{\mathrm{2}}_{jk}}+\nonumber\right.
\\&\left.{Q^{\mathrm{2}}_{il} \delta _{jk}}-
\frac{61}{30}Q^{\mathrm{2}}_{ik} \delta _{jl}+
{Q^{\mathrm{2}}_{ij} \delta _{kl}}\right)+\frac{ \Delta T^{Q^\mathrm{4}} }{ \Delta T^{Q^\mathrm{0}}}\frac{{Q^\mathrm{4}}_{ijkl}}{1008}\Big].
\end{align}

Consider a case where the particles are all oriented in a given direction and their orientation is fixed, due for example to an external field. In addition to a contribution to the viscosity coming from the stresslet, there will be a correction coming from the torque, which is required to keep the particles from rotating with the external flow, $\boldsymbol\Omega = 0 $. These additional contributions come from the resistance tensors connecting the shear rate and the angular velocity to the torque applied to the particles. They are given by
\begin{eqnarray}
\widetilde{\mathcal{R}}_{ijk}^{\mathrm{TE}}&=&
\frac{1}{20}\eta^\infty v^\mathrm{p} \frac{ {\Delta T}^{Q^\mathrm{2}} }{{\Delta T}^{Q^\mathrm{0}}}\left(
Q^\mathrm{2}_{kl}\varepsilon_{ijl}+ Q^\mathrm{2}_{jl}\varepsilon_{ikl}\right)\\
\widetilde{\mathcal{R}}_{ij}^{\mathrm{T\Omega}}&=&\frac{9}{2}\eta^\infty v^\mathrm{p}\left(\delta_{ij}-\frac{1}{15}\frac{ {\Delta T}^{Q^\mathrm{2}} }{{\Delta T}^{Q^\mathrm{0}}}Q^\mathrm{2}_{ij}\right)\label{eq:resistance_torque} \\
{\mathcal{R}}_{ij}^{\mathrm{T\Omega}}&=& 6\eta^\infty v^\mathrm{p} \delta_{ij},\label{eq:resistance_TOmega}
\end{eqnarray} 
where the particle's volume $v^\mathrm{p}=\frac{4\pi a^3}{3}.$
%The resistance tensors in Eqs.~\ref{eq:resistance_stresslet}--\ref{eq:resistance_torque} extend results in Ref.~\cite{opp16}, were
%the relations between force/torque and velocity/angular velocity were given. There, the resistance tensor dependend only on the heat monopole, 
%dipole and quadrople. Here, we see that the stresslet relates to the shear-rate also by the hexadecapole, but no higher orders appear. 
Note that Eq.~\ref{eq:resistance_TOmega} is simply the resistance of a ``cold" fixed particle as coming from Eq.~\ref{eq:resistance_cold}. 

We consider a homogeneous distribution of particles. If the orientations of the particles are uniformly distributed, such as when the particles are dilute and Brownian, there will only be a contribution coming from the heat monopole. All higher-order corrections do not contribute after angular integration. 
In the non-Brownian case, initially oriented particles will stay oriented and contributions from higher moments will appear. We also present corrections to the stress tensor in cases where the particles are all oriented due to an external field. If the orientation is fixed, the particles experience a torque, which will also contribute to the viscosity. 

Assume the fluid is undergoing simple shear, ${\bf u}^{\infty} = y\hat{x}$, such that the shear-rate tensor and vorticity vector are given by Eq.~\ref{eq:simple_shear} with $\dot{\gamma} = 1$.
% \cref{eq:resistance_stresslet} allows us to write the bulk stress for the particles' contribution $\boldsymbol{\Sigma}^{\mathrm{p}}$ as 
% \begin{equation}
%     \boldsymbol{\Sigma}^\mathrm{p}=\boldsymbol{\eta}^\mathrm{p}:{\bf E}^{12}.
% \end{equation}
Now, let us write particles' contribution to the bulk stress, $\boldsymbol{\Sigma}^p$.
First, for uniformly heated particles such that only the heat monopole $Q^0$ exists, the particles' contribution to the bulk-stress is
\begin{equation}
    \boldsymbol{\Sigma}^{{Q^\mathrm{0}}}=\left(\frac{5}{2}-\frac{ 31 }{ 144 }\xi\right)2 {\bf E}^{\infty}
\end{equation}
and the correction to the effective viscosity is
\begin{equation}
    \frac{\eta^{\rm p}}{\eta^{\infty}} = \varphi\left(\frac{5}{2} - \frac{31}{144}\xi \right).
\end{equation}
When there is an external torque orienting the particles, there is an additional contribution to the stress,
\begin{equation}
    \boldsymbol{\Sigma}^{{Q^\mathrm{0}}}=\left(\frac{5}{2}-\frac{ 31 }{ 144 }\xi\right){\bf E}^{\infty}+\frac{3}{2}\boldsymbol{\varepsilon}\left(1-\frac{3}{4}\xi\right)\boldsymbol{\Omega}^{\infty}.
    \end{equation}

\begin{figure}[h!]
\centering\includegraphics[width=0.7\linewidth]{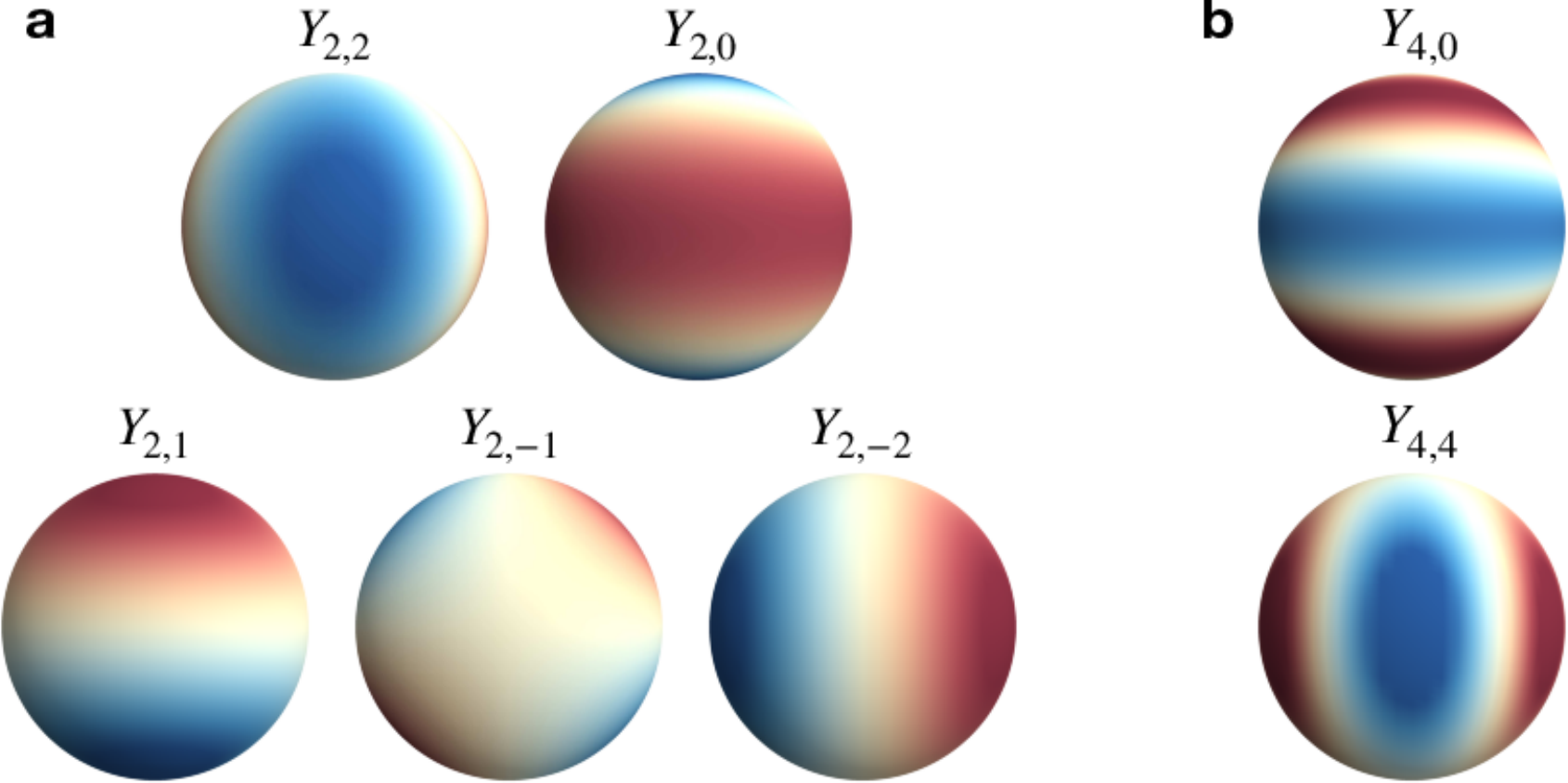}
    \caption{(a) Quadrupole heat moment as spanned by the real spherical harmonics with  $l=2$. Only the first line contributes to the effective viscosity. (b) Hexadecapole heat moment as spanned by real spherical harmonics $(l=4, m=0,4)$. Only components that contribute to the stresslet are presented.}
    \label{figure:y2y4}
\end{figure}

For a complex heat distribution (see \cref{fig:heat_moments_decomp}), we can decompose the heat distribution over the particles' surface by the heat multipoles as described in \cref{eq:eqT}, and represent them using a linear combination of real spherical harmonics, $Y_{l,m}({\bf r})$
\cite{hess2015,edmonds1957angular}.
All odd orders (such as dipole and octopole), do not contribute to the effective viscosity, as well as orders higher than the hexadecapole $Q^{\mathrm{4}}$. Projecting a general heat moment on the different spherical harmonics allows us to identify which terms contribute to the stress tensor. For example, $Q^{l,Y_{l,m}}$ corresponds to the $Y_{l,m}$ component of the $Q^{l}$ heat multipole. The contribution of the quadrupole to the bulk stress is given by
\begin{equation}
\begin{aligned}
  \boldsymbol{\Sigma}^{Q^2} =&\frac{1}{1680}\frac{ {\Delta T}^{Q^\mathrm{2}} }{{\Delta T}^{Q^\mathrm{0}}}
  \begin{bmatrix}
    -2Q^{2, Y_{2,-2}} & 62Q^{2, Y_{2,0}}&-31Q^{2 ,Y_{2,-1}}\\
    62Q^{2, Y_{2,0}}&-2Q^{2, Y_{2,-2}}&-31Q^{2, Y_{2,1}}\\
    -31Q^{2, Y_{2,-1}}&-31Q^{2, Y_{2,1}}&60Q^{2, Y_{2,-2}}
    \end{bmatrix}.
\end{aligned}
\end{equation}
Notice that the stress tensor can no longer be mapped to the strain-rate tensor by a scalar field. Therefore the viscosity in such a case is tensorial. 
The contribution to the scalar effective viscosity, coming from $\Sigma_{12}$, is given by
\begin{equation}
    \frac{\eta^{\rm p}}{\eta^{\infty}} = \varphi\left(\frac{5}{2} - \xi\frac{31}{840}\frac{ {\Delta T}^{Q^\mathrm{2}} }{{\Delta T}^{Q^\mathrm{0}}}Q^{2,Y_{2,0}} \right).%60
\end{equation}
Figure~\ref{figure:y2y4}a gives the decomposition of the quadrupole to the different real spherical harmonics, including $Y_{2,0}$, which contributes to the effective viscosity coming from a hot stresslet. 

For the hexadecapole, the bulk stress tensor is given by
\begin{equation}
\begin{aligned}
  \boldsymbol{\Sigma}^{Q^4} =&\frac{1}{1008}\frac{ {\Delta T}^{Q^\mathrm{4}} }{{\Delta T}^{Q^\mathrm{0}}}
  \begin{bmatrix}
    Q^{4, Y_{4,-4}}-Q^{4, Y_{4,-2}} & 
    6 \left(Q^{4, Y_{4,0}}-Q^{4, Y_{4,4}}\right)&
    3\left(Q^{4 ,Y_{4,-3}}-Q^{4 ,Y_{4,-1}}\right)\\
    6 \left(Q^{4, Y_{4,0}}-Q^{4, Y_{4,4}}\right)& 
    3Q^{4 ,Y_{4,-2}}-Q^{4 ,Y_{4,-4}}&
    -3\left(Q^{4 ,Y_{4,1}}+Q^{4 ,Y_{4,3}}\right)\\
    3\left(Q^{4 ,Y_{4,-3}}-Q^{4 ,Y_{4,-1}}\right) &
    -3\left(Q^{4 ,Y_{4,1}}+Q^{4 ,Y_{4,3}}\right) &
    -6Q^{4 ,Y_{4,-2}}
    \end{bmatrix},
\end{aligned}
\end{equation}
and the correction to the effective viscosity, $\Sigma_{12}$, is given by
\begin{equation}
    \frac{\eta^{\rm p}}{\eta^{\infty}} = \varphi\left(\frac{5}{2} - \xi\frac{1}{168}\frac{ {\Delta T}^{Q^\mathrm{4}} }{{\Delta T}^{Q^\mathrm{0}}}\left(Q^{4, Y_{4,0}}-Q^{4, Y_{4,4}} \right)\right).%60
\end{equation}
Figure~\ref{figure:y2y4}b shows the real spherical harmonic components of the hexadecapole that contribute to the stresslet.

If there is an external torque fixing the particles at a given orientation, angular momentum is no longer conserved and therefore the stress tensor is not necessarily symmetric. The quadrupolar contribution to the stress tensor, ${\Sigma}^{Q^2}$, will include additional terms as given by \cref{eq:paticles_bulk_stress}. The first contribution is related to the effect of the background vorticity on the stresslet. It is given by
\begin{equation}
\begin{aligned}
    \boldsymbol{\Sigma}^{Q^2, \mathcal{S}\Omega}
    =& \frac{1}{40}\frac{ {\Delta T}^{Q^\mathrm{2}} }{{\Delta T}^{Q^\mathrm{0}}}\begin{bmatrix}
2Q^{2, Y_{2,-2}}&-2Q^{2,Y_{2,2}}&Q^{2, Y_{2,-1}}\\
-2Q^{2, Y_{2,2}}&-2Q^{2, Y_{2,-2}}&-Q^{2, Y_{2,1}}\\
Q^{2, Y_{2,-1}}&-Q^{2, Y_{2,1}}&0
    \end{bmatrix}.
\end{aligned}
\end{equation}
The second contribution comes from the effect of the background vorticity on the torque. It is given by
\begin{equation}
\begin{aligned}
    \boldsymbol{\Sigma}^{Q^2, {\rm T} \Omega}
    =& \frac{3}{40}\frac{ {\Delta T}^{Q^\mathrm{2}} }{{\Delta T}^{Q^\mathrm{0}}}\begin{bmatrix}
0&-2Q^{2, Y_{2,0}}&Q^{2, Y_{2,-1}}\\
2Q^{2, Y_{2,0}}&0&-Q^{2, Y_{2,1}}\\
-Q^{2, Y_{2,-1}}&Q^{2, Y_{2,1}}&0
    \end{bmatrix}.
\end{aligned}
\end{equation}
The third contribution comes from the effect of the shear-rate tensor on the torque and is given by 
\begin{equation}
\begin{aligned} 
    \boldsymbol{\Sigma}^{Q^2, {\rm T}{\rm E}}
    = &\frac{1}{40}\frac{ {\Delta T}^{Q^\mathrm{2}} }{{\Delta T}^{Q^\mathrm{0}}}\begin{bmatrix}
0&-2Q^{2, Y_{2,2}}&Q^{2, Y_{2,-1}}\\
2Q^{2, Y_{2,2}}&0&Q^{2, Y_{2,1}}\\
-Q^{2, Y_{2,-1}}&-Q^{2, Y_{2,1}}&0
    \end{bmatrix}.
\end{aligned}
\end{equation}
The new terms that appear in the stress tensor obey certain symmetries
\begin{align}
    &\Sigma^{\mathcal{S}{\rm E}}_{ijkl}=\Sigma^{\mathcal{S}{\rm E}}_{ijlk}=\Sigma^{\mathcal{S}{\rm E}}_{ijlk}
    ,\label{eq:stresslet_shear_exapmle}\\ 
    &\Sigma^{\rm TE}_{ijkl}=-\Sigma^{\rm TE}_{ijlk}=\Sigma^{\rm TE}_{jikl}
    ,\label{eq:torque_shear_exapmle} \\ 
    &\Sigma^{\rm T\Omega}_{ijkl}=-\Sigma^{\rm T\Omega}_{ijlk}=-\Sigma^{\rm T\Omega}_{jikl}
    ,\label{eq:torque_angvel_exapmle} \\ 
    &\Sigma^{\mathcal{S}{\Omega}}_{ijkl}=\Sigma^{\mathcal{S}{\Omega}}_{ijlk}=-\Sigma^{\mathcal{S}{\Omega}}_{ijlk}
    ,\label{eq:stresslet_angvel_exapmle}
\end{align}
where \cref{eq:torque_shear_exapmle,eq:stresslet_angvel_exapmle} represent an odd viscosity property of the suspension.

In total, the correction to the effective viscosity from the particles is given by
\begin{equation}
\label{eq:totalParticleViscosity}
\begin{aligned}
   \frac{1}{\varphi} \frac{\eta^{\rm p}}{\eta^{\infty}} =&
  \underbrace{\frac{5}{2}}_\text{Einstein}+ \ \underbrace{\frac{3}{2}}_\text{Cold torque} \ \ \ 
    % \underbrace{-\left(\frac{31}{144}\xi +\frac{9}{8}\xi\right)}_\text{Monopole}\\
    \underbrace{-\frac{193}{144}\xi}_\text{Monopole}\\
    &- \xi\underbrace{\frac{ {\Delta T}^{Q^\mathrm{2}} }{{\Delta T}^{Q^\mathrm{0}}}\left(\frac{1}{10}Q^{2,Y_{2,2}}+\frac{95}{840}Q^{2,Y_{2,0}}
    \right)}_\text{Quadrupole}- \underbrace{\xi\frac{1}{168}\frac{ {\Delta T}^{Q^\mathrm{4}} }{{\Delta T}^{Q^\mathrm{0}}}\left(Q^{4, Y_{4,0}}-Q^{4, Y_{4,4}}\right)}_\text{Hexadecapole}.%60
\end{aligned}
\end{equation}
In addition, a consequence of a non-isotropic fluid is that the fluid is non-Newtonian at $O(\xi\phi)$, as there can be normal-stress differences for certain heat distributions
\begin{eqnarray}    
    &N_1\equiv\langle\Sigma_{11}\rangle-\langle\Sigma_{22}\rangle=-{\varphi}
    \xi\left(\frac{1}{10}\frac{ {\Delta T}^{Q^\mathrm{2}} }{{\Delta T}^{Q^\mathrm{0}}}Q^{2,Y_{2,-2}}+\frac{2}{1008}
                {\frac{ {\Delta T}^{Q^\mathrm{4}} }{{\Delta T}^{Q^\mathrm{0}}}\left(Q^{2,Y_{4,-2}}+Q^{2,Y_{4,-4}}\right)}\right),\\
    &N_2\equiv\langle\Sigma_{22}\rangle-\langle\Sigma_{33}\rangle=-{\varphi}
    \xi\left(-\frac{73}{840}\frac{ {\Delta T}^{Q^\mathrm{2}} }{{\Delta T}^{Q^\mathrm{0}}}Q^{2,Y_{2,-2}}+\frac{1}{1008}
                {\frac{ {\Delta T}^{Q^\mathrm{4}} }{{\Delta T}^{Q^\mathrm{0}}}\left(3Q^{2,Y_{4,-2}}-Q^{2,Y_{4,-4}}\right)}\right).
\end{eqnarray}

Lastly, let us present another consequence of the tensorial viscosity by using a different simple shear, such as is caused by the flow
\begin{equation}
    {\bf u}^\infty=x\hat{y},
\end{equation}
which gives the background strain-rate tensor and vorticity vector
\begin{equation}
\label{eq:simple_shear2}
\begin{aligned}
        {\bf E}^{\infty}=\frac{\dot\gamma}{2}\begin{bmatrix}
        0&1&0\\1&0&0\\0&0&0
    \end{bmatrix}
    &\quad ; \quad \boldsymbol{\Omega}^{\infty}=-\frac{\dot\gamma}{2}\begin{bmatrix}
        0\\0\\1
    \end{bmatrix}.
\end{aligned}
\end{equation}
In this case, the scalar effective viscosity is given by ${ \eta}^{\mathrm{p}}=\Sigma_{21}/\dot\gamma$. Note that the sole difference is in the sign of the vorticity vector, yet the effective viscosity, calculated from \cref{eq:stresslet_shear_exapmle,eq:stresslet_angvel_exapmle,eq:torque_angvel_exapmle,eq:torque_shear_exapmle}, gives a scalar effective viscosity which is different from that of Eq.~\ref{eq:totalParticleViscosity}
\begin{equation}
\begin{aligned}
   \frac{1}{\varphi} \frac{\eta^{\rm p}}{\eta^{\infty}} =&
  \underbrace{\frac{5}{2}}_\text{Einstein}+\underbrace{\frac{3}{2}}_\text{Cold torque} \ \
    % \underbrace{-\left(\frac{31}{144}\xi +\frac{9}{8}\xi\right)}_\text{Monopole}\\
    \underbrace{-\frac{193}{144}\xi}_\text{Monopole}\\
    &- \xi\underbrace{\frac{ {\Delta T}^{Q^\mathrm{2}} }{{\Delta T}^{Q^\mathrm{0}}}\left(\frac{1}{10}Q^{2,Y_{2,2}}-\frac{95}{840}Q^{2,Y_{2,0}}
    \right)}_\text{Quadrupole}- \underbrace{\xi\frac{1}{168}\frac{ {\Delta T}^{Q^\mathrm{4}} }{{\Delta T}^{Q^\mathrm{0}}}\left(Q^{4, Y_{4,0}}-Q^{4, Y_{4,4}}\right)}_\text{Hexadecapole}.%60
\end{aligned}
\end{equation}
We can conclude that not only is the suspension anisotropic due to the coupling to the particles' heat orientation $(Q^{2, Y_{2,m}},Q^{4, Y_{4,m}})$, but also that it can be presented more accurately by a tensorial viscosity, \cref{eq:eff_visc_tensor}, as for example, $\eta^\mathrm{eff}_{2121}\neq \eta^\mathrm{eff}_{1212}$.

\section{Conclusions}
To conclude, in this paper, we calculated the effective stress tensor and effective viscosity of a suspension of hot particles. The particles are assumed to be heated (for example, by an external field), and the temperature distribution over the particles' surface is given by $\mathcal{T}({\bf r})$. The temperature affects the local
viscosity. We assumed a small temperature difference defined by a small-parameter, $\xi$. 
In order to calculate the correction to the viscosity, we used the extended Lorentz reciprocal theorem. First, we used it to calculate an extensive correction to the viscosity coming from the heated fluid. This correction depends on system size --- for a given concentration of hot particles, a larger system is hotter than a smaller one, therefore its viscosity will be lower in comparison. The second correction comes from the particles themselves --- each particle is force-free and, therefore, changes the stress distribution by the force dipole. In cases where there is no external torque, the particles only affect the flow by the symmetric part of the force dipole --- the stresslet. The stresslet is modified by the temperature distribution on the particle. For spherical particles, and if the particles are free to rotate, the only correction comes from the heat monopole. Higher-order multipoles do not contribute. The correction to the stresslet is negative. If the particles are non-Brownian and are all oriented in the same direction, there may be additional contributions to the stresslet from the quadrupole and the hexadecapole. The fluid is no longer isotropic, and the viscosity, in principle, has a tensorial form as the relationship between external strain rate and the fluid's stress. In such a case, the fluid is also non-Newtonian as there are normal stress differences. If the particles' orientation is fixed due to an external field, there is a torque acting on them, prohibiting rotation. In such a case, angular momentum is no longer conserved, and the stress tensor is not symmetric as there are contributions coming from the vorticity field and the torque. This latter case can be thought of as a fluid with odd viscosity.

\bibliography{hot_par}

\end{document}